\documentclass[a4paper,11pt]{article}
\usepackage{pos}
\usepackage{physics}

\title{Riemannian manifold hybrid Monte Carlo in lattice QCD}

\author*[a]{Tuan Nguyen}
\author[b]{Peter Boyle}
\author[a]{Norman Christ}
\author[c]{Yong-Chull Jang}
\author[c]{Chulwoo Jung}

\affiliation[a]{Department of Physics, Columbia University,\\
New York, NY 10027, USA}

\affiliation[b]{School of Physics and Astronomy, University of Edinburgh,\\
Edinburgh EH9 3JZ, United Kingdom}

\affiliation[c]{Physics Department, Brookhaven National Laboratory,\\
Upton, NY 11973, USA}

\emailAdd{thn2112@columbia.edu}
\emailAdd{paboyle@ph.ed.ac.uk}
\emailAdd{nhc@phys.columbia.edu}
\emailAdd{ypj@bnl.gov}
\emailAdd{chulwoo@bnl.gov}

\abstract{Critical slowing down presents a critical obstacle to lattice QCD calculation at the smaller lattice spacings made possible by Exascale computers. Inspired by the concept of Fourier acceleration, we study a version of the Riemannian Manifold HMC (RMHMC) algorithm in which the canonical mass term of the HMC algorithm is replaced by a rational function of the SU(3) gauge covariant Laplacian. We have developed a suite of tools using Chebyshev filters based on the SU(3) gauge covariant Laplacian that provides the power spectra of both the gauge and fermion forces and determines the spectral dependence of the resulting RMHMC evolution of long- and short-distance QCD observables. These tools can be used to optimize the RMHMC mass term and to monitor the resulting acceleration mode-wise.}

\FullConference{%
 The 38th International Symposium on Lattice Field Theory, LATTICE2021
  26th-30th July, 2021
  Zoom/Gather@Massachusetts Institute of Technology
}

\begin{document}
\maketitle

\section{Introduction}

Overcoming critical slowing down is a crucial objective of modern lattice computations. The problem of critical slowing down stems from the increasing range of distance or energy scales that must be accommodated in the usual Markov-chain gauge-evolution algorithms as the continuum limit is approached.  This problem is made worse by the inherent non-renormalizability of the favored Hybrid Monte Carlo (HMC) algorithm \cite{luscher_non-renormalizability_2011}.  For free fields, we expect for fixed physical lattice size but with decreasing lattice spacing $a$ that non-local observables will decorrelate after integrating over a physical trajectory length \cite{Kennedy:2000ju}.   Because of the higher frequency modes the trajectory step size must scale with the lattice spacing $a$, meaning the naive autocorrelation time scales as $a^{-1}$. However, L\"uscher and Schaefer present evidence  that the true autocorrelation time of these observables scales as $a^{-2}$ \cite{Luscher2011} and suggest that this Langevin-like behavior results from  the non-renormalizability of the HMC algorithm.

The goal of Fourier acceleration is to modify the kinetic energy term in the molecular dynamics Hamiltonian with a coordinate-dependent mass term such that the low modes move at a higher MD velocity. In certain systems, a coordinate-dependent mass term can be chosen such that all modes move at the same velocity and critical slowing down is eliminated. For example, consider a lattice pure $U(1)$ gauge theory in the continuum limit. Substituting the gauge links with gauge fields via $U_\mu(x) = \exp(i A_\mu(x) / \sqrt{2\beta})$ and changing the MD mass term to be $\mathcal{M} = (1-\kappa) - \kappa\partial^2$ the MD Hamiltonian becomes
\begin{equation}
    \mathcal{H} = \frac{1}{2} P_\mu \qty[(1-\kappa) - \kappa\partial^2]^{-1} P_\mu + \frac{1}{2} A_\mu (\partial_\mu\partial_\nu - \delta_{\mu\nu}\partial^2) A_\nu + \mathcal{O}(1/\beta)\,.
\end{equation}
Since the gauge fields are abelian, MD updates on the gauge links commute with gauge transformations, and thus we are free to fix the system to Lorenz gauge. Therefore the gauge modes are simply plane waves that evolve as $\dv*[2]{\tilde{A}_\mu(k)}{\tau} = -\omega^2(k) \tilde{A}_\mu(k)$ with characteristic frequency
\begin{equation}
    \omega^2(k) = \frac{k^2}{1 - \kappa + \kappa k^2} \,.
\end{equation}
In the limit $\kappa \to 1$, all modes move with the same velocity and the modified HMC is fully accelerated. \cite{Duane1986} This same choice of mass term may also accelerate $\phi^4$ theories in the perturbative limit since the eigenmodes of the system are still approximately plane waves, meaning one can tune the mode-dependence of the mass term to achieve approximate Fourier acceleration. \cite{christ2018fourier}

The situation is more complicated when we try to accelerate the HMC for lattice QCD since the theory is non-abelian and non-perturbative. We can parameterize the coordinate dependent mass term as a function of the SU(3) gauge-covariant Laplace operator in order to maintain the gauge invariance of the MD Hamiltonian. However, in order to use the eigenmodes of the covariant Laplace operator to tune the mass term for Fourier acceleration we must show that the eigenmodes of the theory have sufficient overlap with the covariant Laplace eigenmodes. As we will show, when projected onto the subspace of Laplace eigenvectors whose eigenvalues lie in a narrow interval $[\lambda - \Delta \lambda/2, \lambda + \Delta \lambda/2]$ the size of the HMC force that results from the gauge action is nearly linear with respect to $\lambda$. Additionally, we will show that an extension of Fourier acceleration for non-abelian pure gauge theories, the Riemannian Manifold HMC (RMHMC) algorithm, is capable of moving certain long-distance observables faster than the naive HMC, indicating some level of Fourier acceleration of the theory.

\section{The HMC Gauge Force Power Spectrum}

\begin{figure}[h]
\centering
\includegraphics[width=0.5\textwidth]{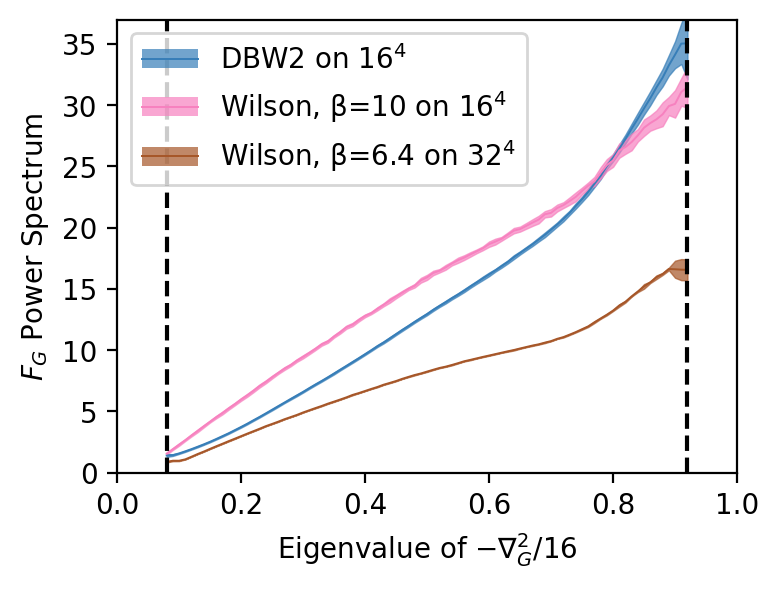}
\caption{The power spectrum per mode of the gauge force for various lattice actions is plotted with respect to the eigenvalues corresponding to each eigenmode of the normalized covariant Laplace operator. The inverse lattice spacings of each action are $a^{-1} = 2\ \text{GeV}$ for the DBW2 action, $a^{-1} = 270\ \text{GeV}$ for the Wilson $\beta = 10$ action, and $a^{-1} = 3.6\ \text{GeV}$ for the Wilson $\beta = 6.4$ action. We cut off the plot for very low and very high eigenvalues of the normalized covariant Laplace operator because there are too few covariant Laplace modes corresponding to these eigenvalues to accurately measure the power spectrum.} \label{fig:fgauge}
\end{figure}

In order to show that the eigenmodes of the covariant Laplace operator can be used to Fourier accelerate the HMC algorithm, we calculate the spectral dependence of the naive HMC gauge force per mode with respect to the eigenvalues of the normalized covariant Laplace operator $-\laplacian_G/(4d)$ where $\laplacian_G$ is defined in the adjoint representation as
\begin{equation}
    \laplacian_G\phi(x) = \sum_\mu^d\qty[U_\mu(x)\phi(x+\mu)U_\mu^\dagger(x) + U_\mu(x-\mu)^\dagger\phi(x-\mu)U_\mu^\dagger(x-\mu) - 2\phi(x)] \,.
\end{equation}
We note that the covariant Laplace operator explicitly depends on the gauge links, which can cause subtle complications that will be addressed later when we present the Riemannian Manifold HMC algorithm.

We define a bandpass filter $\mathcal{B}(\lambda) \approx P_{\lambda,\Delta\lambda}\qty[-\laplacian_G/16]$ that approximates a projection operator onto a narrow range of eigenvalues of $-\laplacian_G/16$ with a sum of Chebyshev polynomials: $\mathcal{B}(\lambda) = \sum c_i T_i[-\laplacian_G/16]$. Using this bandpass filter, we project the gauge force $F_G = -\pdv*{S[U]}{U_\mu(x)}$ (where $\pdv*{U_\mu}$ represents the conventional derivative with respect to the group parameters resulting in the left multiplication of the link matrix $U_\mu$ by the group generators) onto a narrow interval of eigenvalues near $\lambda$ to extract the contributions from covariant Laplace eigenmodes corresponding to those eigenvalues. We calculate the power spectrum of the gauge force per mode as
\begin{equation}
    \mel{F_G}{\mathcal{B}(\lambda)}{F_G} \Big/ \Tr[\mathcal{B}(\lambda)] \,.
\end{equation}

The gauge force power spectrum per mode is plotted in Figure \ref{fig:fgauge} for various lattice actions. The power spectrum is a relative measure of how quickly an eigenmode of the covariant Laplace operator would move after integrating a small trajectory time step. For the pure $U(1)$ gauge theory for example, the power spectrum is a linear function of the wavenumber squared. From Figure \ref{fig:fgauge} we see that power spectrum is surprisingly nearly linear with respect to the eigenvalue of $-\laplacian_G/16$ for all actions, which confirms that the eigenmodes of the gauge action are approximately covariant Laplace eigenmodes.

\section{The RMHMC Algorithm}

The Riemannian manifold HMC (RMHMC) is a class of Fourier accelerated HMC algorithms where we add a coordinate dependent mass term that depends on the covariant Laplace operator. The full Hamiltonian is given by
\begin{equation}
    \mathcal{H} = \dfrac{1}{2}\Tr\qty{P_\mu^\dagger(x)\qty[\mathcal{M}\qty(-\laplacian_G/16)]^{-1}P_\mu(x)} + S\qty[U_\mu(x)] + \log\abs{\mathcal{M}}
\end{equation}
The addition of the $\log\abs{\mathcal{M}}$ term is required in order to cancel the additional force $F_M = -\Tr\qty{P^\dagger[\pdv*{\mathcal{M}^{-1}}{U_\mu(x)}]P}$ caused by the dependence of the mass term on the gauge links. This additional term in the Hamiltonian will be realized by adding auxiliary fields and auxiliary conjugate momenta. A further complication is the nonseparability of the Hamiltonian, which requires an implicit leap-frog integration scheme.\cite{Girolami2011}

In Duane et al., they consider a simple mass term that is linear in the gauge-invariant Laplacian. \cite{Duane1986} Our initial studies of this simple mass term show little if any reduction in autocorrelation. We therefore consider more general sums of rational functions of the covariant Laplace operator to parameterize our mass term:
\begin{equation}
    \mathcal{M} = \qty[c + \sum_{i=1}^N \dfrac{a_{0,i}+a_{1,i}\qty(-\laplacian_G/16)}{b_{0,i}+b_{1,i}\qty(-\laplacian_G/16)+\qty(-\laplacian_G/16)^2}]^{-2} \,.
\end{equation}

\begin{figure}[h]
\centering
\includegraphics[width=0.5\textwidth]{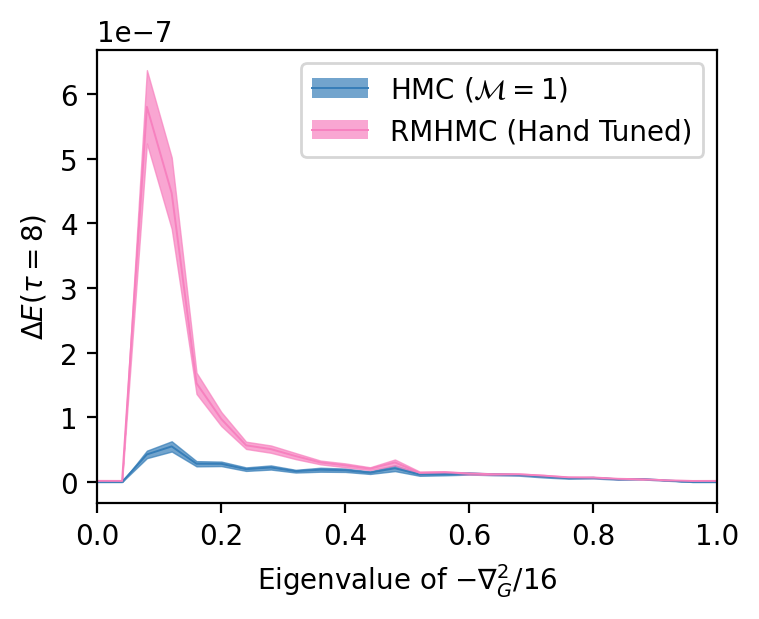}
\centering
\caption{
The change in Wilson flowed energy (flow time of 8) after a single HMC leapfrog integration step of $10^{-4}$ MD time units with injected conjugate momenta projected onto specific covariant Laplace eigenmodes. All algorithms have the largest change in Wilson flowed energy after injecting low modes.
All calculations are done with a DBW2 action with $a^{-1} = 2\ \text{GeV}$ on a $16^4$ lattice.} \label{fig:E8}
\end{figure}

For our current study we did not pursue a full-length autocorrelation study between the naive HMC and the RMHMC. Instead, we investigate whether or not the RMHMC is able to move long-distance observables further than the naive HMC, allowing us to more quickly probe whether we have indeed Fourier accelerated the theory. We chose to study changes in the Wilson flowed energy between the two algorithms, since Wilson flowed observables are sensitive to length-scales on the order of the square root of the flow time. \cite{Luscher2010}

As a preliminary test of the correlation between the Laplace eigenmodes with small eigenvalues and the long-distance physics measured by the Wilson flowed energy at large flow time, we examine how sensitive the Wilson flowed energy is to perturbations to the gauge links by specific covariant Laplace modes. We compare the naive HMC algorithm to an RMHMC algorithm with a mass term that is hand-tuned to approximate the spectral dependence of the gauge force term while remaining easy to invert. For the Wilson action, we consider an additional RMHMC mass term that is fitted as a sum of many rational functions to match the spectral dependence of the gauge force, which should theoretically give optimal acceleration.

For each algorithm, we generate random conjugate momenta as if initiating an HMC trajectory with a heatbath, but then use the bandpass filters from before to project these momenta onto a narrow range of covariant Laplace eigenvalues. Since the HMC heatbath depends on the mass term for the generation of random conjugate momenta, the RMHMC algorithm will generate larger momenta corresponding to low covariant Laplace eigenmodes. We then inject these filtered momenta into a short HMC trajectory and measure the change in the Wilson flowed energy.
As shown in Figure \ref{fig:E8}, the Wilson flowed energy is pushed further by the RMHMC algorithms compared to the naive HMC during a single, short HMC integration step after injecting momenta corresponding to low modes, which indicates that the covariant Laplace eigenmodes with small eigenvalues are highly correlated with long-distance scales.

\begin{figure}[h]
\centering
\begin{minipage}{0.48\textwidth}
\includegraphics[width=\textwidth]{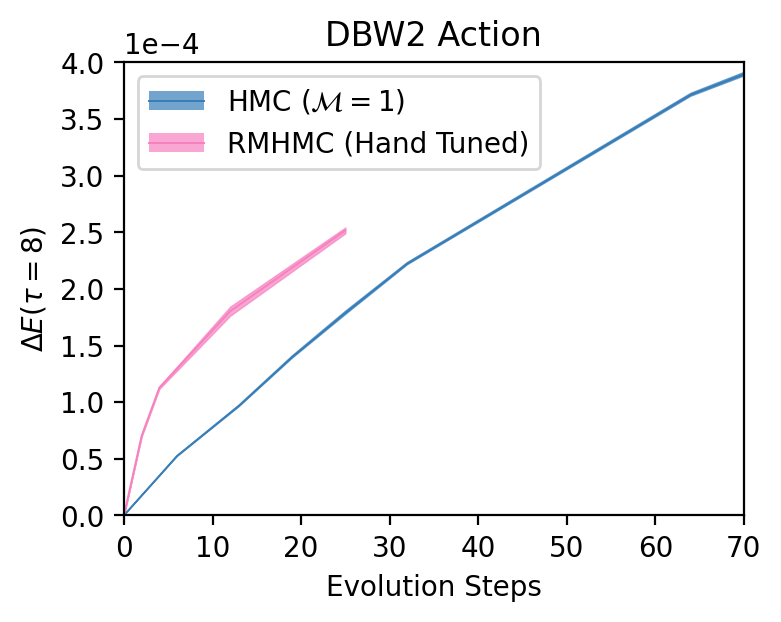}
\end{minipage}\hfill
\begin{minipage}{0.5\textwidth}
\includegraphics[width=\textwidth]{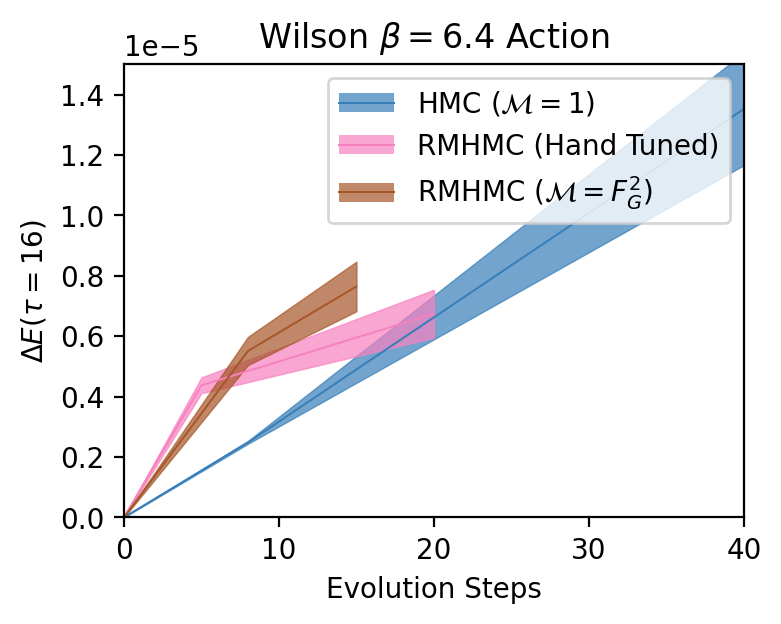}
\end{minipage}\hfill
\centering
\caption{The change in Wilson flowed energy from that of the initial configuration after a single HMC or RMHMC trajectory versus the number of MD integration steps in that trajectory. The left plot is run with a DBW2 action with $a^{-1} = 2\ \text{GeV}$ on a $16^4$ lattice, while the right plot is run with a Wilson action with $\beta = 6.4$ and $a^{-1} = 3.6\ \text{GeV}$ on a $32^4$ lattice. Due to working on a larger lattice for $\beta = 6.4$, we double the flow time to 16 for the right plot.} \label{fig:EMD}
\end{figure}

Finally, we plot the change in the Wilson flowed energy after a single trajectory as a function of the number of MD integration steps in that trajectory for each algorithm to see if the RMHMC algorithms indeed change the long-distance stucture of the gauge configurations faster than the naive HMC. For each of our algorithms, we tune the MD trajectory step size so that all algorithms give us similar accept/reject probabilities. Our results are shown in Figure \ref{fig:EMD} for two actions, a DBW2 action with $a^{-1} = 2\ \text{GeV}$ on a $16^4$ lattice and a Wilson action with $\beta = 6.4$ and $a^{-1} = 3.6\ \text{GeV}$ on a $32^4$ lattice. For both actions, we see that the RMHMC algorithms are able to move the Wilson flowed energy about twice as fast in the first few MD evolution steps than the naive HMC, before slowing down to about the same rate of change as the naive HMC.
Thus, we conclude that we are able to accelerate the low modes of the theory for the actions considered and that the RMHMC is effective at moving long-distance observables faster than the HMC at similar computation cost, {\it i.e.} with a similar number of integration steps at fixed acceptance.

\section{Conclusion and Next Steps}

\begin{figure}[h!]
\centering
\includegraphics[width=0.5\textwidth]{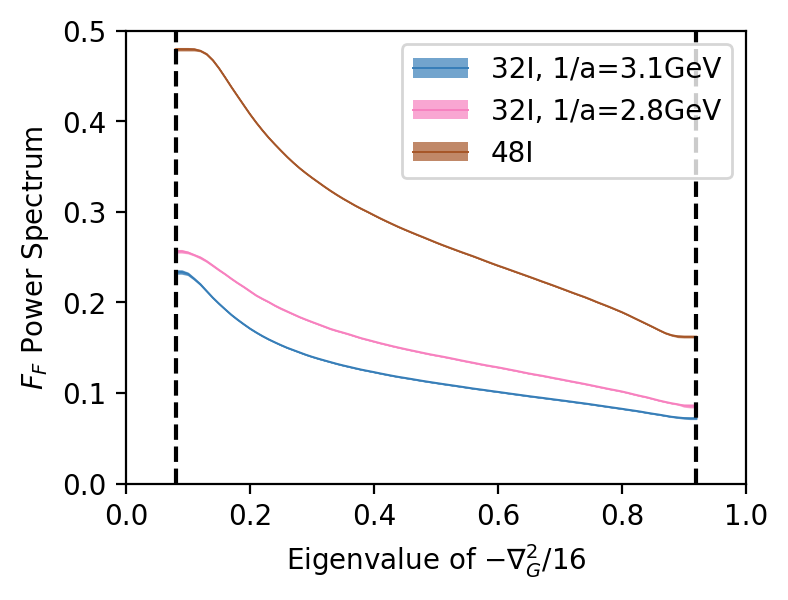}
\caption{The power spectrum per mode of the fermion force for various Iwasaki actions of different lattice sizes and coupling constants. Axes are the same as in Figure \ref{fig:fgauge}} \label{fig:fferm}
\end{figure}

In this work, we propose the RMHMC class of algorithms for Fourier accelerating the HMC algorithm by modifying the mass term in the MD Hamiltonian such that the low modes of the theory can be moved more quickly. By measuring the Wilson flowed energy, an observable that is sensitive to the long-distance structure of the gauge configurations, we have shown that the RMHMC algorithm is able to move these low modes in fewer evolutions steps than the naive HMC. Ultimately, we are interested in using the RMHMC with fermions. Applying the same techniques as before to measure the power spectrum of the fermion force per mode, we show in Figure \ref{fig:fferm} that the fermion force has a milder spectral dependence than the gauge force, and it already moves low modes further than high modes. This is unsurprising since the fermion determinant is a highly non-local operator. We expect that the RMHMC can be applied to a hierarchical integration scheme where we separate the fermion force and the gauge force onto different integration levels, where we would only need the implicit integration scheme with the lower, gauge-only levels.

In future studies, we are interested in running a full-length study of the autocorrelation time of some long-distance observables to show definitive reduction in autocorrelation time. In addition, we would like to tune and run the RMHMC algorithm with fermions. One potential first step in this direction would be to investigate the spectral dependence of fermion forces from individual Hasenbusch ratios.

\section*{Acknowledgements}

We thank our RBC and UKQCD Collaboration colleagues, in particular Joseph Karpie and Robert Mawhinney, for helpful discussions and ideas.  This research was supported by the Exascale Computing Project (17-SC-20-SC), a collaborative effort of the U.S. Department of Energy Office of Science and the National Nuclear Security Administration.   We are grateful for the computational resources provided by Argonne Leadership Computing Facility, which is a DOE Office of Science User Facility supported under Contract DE-AC02-06CH11357 and the Oak Ridge Leadership Computing Facility at the Oak Ridge National Laboratory, which is supported by the Office of Science of the U.S. Department of Energy under Contract No. DE-AC05-00OR22725.  In addition NHC was supported by U.S. DOE grant \#DE-SC0011941.

\bibliographystyle{JHEP}
\bibliography{refs}

\providecommand{\href}[2]{#2}\begingroup\raggedright\begin{thebibliography}{1}

\bibitem{luscher_non-renormalizability_2011}
M.~Lüscher and S.~Schaefer, \emph{Non-renormalizability of the {HMC}
  algorithm}, \href{https://doi.org/10.1007/JHEP04(2011)104}{\emph{Journal of
  High Energy Physics} {\bfseries 2011} (2011) 104}.

\bibitem{Kennedy:2000ju}
A.D.~Kennedy and B.~Pendleton, \emph{{Cost of the generalized hybrid Monte
  Carlo algorithm for free field theory}},
  \href{https://doi.org/10.1016/S0550-3213(01)00129-8}{\emph{Nucl. Phys. B}
  {\bfseries 607} (2001) 456}
  [\href{https://arxiv.org/abs/hep-lat/0008020}{{\ttfamily hep-lat/0008020}}].

\bibitem{Luscher2011}
M.~Lüscher and S.~Schaefer, \emph{Lattice qcd without topology barriers},
  \href{https://doi.org/10.1007/jhep07(2011)036}{\emph{Journal of High Energy
  Physics} {\bfseries 2011} (2011) }.

\bibitem{Duane1986}
S.~Duane, R.~Kenway, B.J.~Pendleton and D.~Roweth, \emph{Acceleration of gauge
  field dynamics},
  \href{https://doi.org/https://doi.org/10.1016/0370-2693(86)90940-8}{\emph{Physics
  Letters B} {\bfseries 176} (1986) 143}.

\bibitem{christ2018fourier}
N.H.~Christ and E.W.~Wickenden, \emph{Fourier acceleration, the hmc algorithm
  and renormalizability},  2018.

\bibitem{Girolami2011}
M.~Girolami and B.~Calderhead, \emph{Riemann manifold langevin and hamiltonian
  monte carlo methods},
  \href{https://doi.org/https://doi.org/10.1111/j.1467-9868.2010.00765.x}{\emph{Journal
  of the Royal Statistical Society: Series B (Statistical Methodology)}
  {\bfseries 73} (2011) 123}
  [\href{https://arxiv.org/abs/https://rss.onlinelibrary.wiley.com/doi/pdf/10.1111/j.1467-9868.2010.00765.x}{{\ttfamily
  https://rss.onlinelibrary.wiley.com/doi/pdf/10.1111/j.1467-9868.2010.00765.x}}].

\bibitem{Luscher2010}
M.~L{\"u}scher, \emph{Properties and uses of the wilson flow in lattice qcd},
  \href{https://doi.org/10.1007/JHEP08(2010)071}{\emph{Journal of High Energy
  Physics} {\bfseries 2010} (2010) 71}.

\end{thebibliography}\endgroup

\end{document}